\documentclass{article}
\usepackage{spconf}
\usepackage{spconf,amsmath,amsfonts,graphicx}
\usepackage{enumitem}
\usepackage{hyperref}
\usepackage{color}
\usepackage{booktabs, multirow}
\usepackage{subfig}

\graphicspath{ {imgs/} }

\newcommand{\figref}[1]{\figurename~\ref{#1}}
\newcommand{\tabref}[1]{\tablename~\ref{#1}}
\newcommand{\txtsub}[2]{$\text{#1}_{#2}$}
\newcommand{\ptxtsub}[2]{$\text{#1}'_{#2}$}

\title{Toward Expressive Singing Voice Correction:  
On Perceptual Validity of Evaluation Metrics for Vocal Melody Extraction}

\name{Yin-Jyun Luo\thanks{The first and second authors performed the work while at Academia Sinica. The first author is now a PhD student at EECS, Queen Mary University of London, UK; the second author now works at Rayark Inc., Taiwan.} \qquad Yuen-Jen Lin \qquad Li Su}
  
 \address{Institute of Information Science, Academia Sinica, Taiwan}

\begin{document}
\ninept
\maketitle
\begin{abstract}
Singing voice correction (SVC) is an appealing application for amateur singers.
Commercial products automate SVC by snapping pitch contours to equal-tempered scales, which could lead to deadpan modifications.
Together with the neglect of rhythmic errors, extensive manual corrections are still necessary.
In this paper, we present a streamlined system to automate expressive SVC for both pitch and rhythmic errors.
Particularly, we extend a previous work by integrating advanced techniques for singing voice separation (SVS) and vocal melody extraction.
SVC is achieved by temporally aligning the source-target pair, followed by replacing pitch and rhythm of the source with those of the target.
We evaluate the framework by a comparative study for melody extraction which involves both subjective and objective evaluations, whereby we investigate perceptual validity of the standard metrics through the lens of SVC.
The results suggest that the high pitch accuracy obtained by the metrics does not signify good perceptual scores.
\end{abstract}
\begin{keywords}
Singing voice correction, audio synchronization, vocal melody extraction, evaluation
\end{keywords}
\section{Introduction}\label{sec:intro}
Automatic singing voice correction (SVC) has found many applications in the era of digital music, 
ranging from mobile apps, karaoke devices, to music production tools~\cite{salazar2015continuous, cano2000voice, melodyne}.
Commercial products mostly rely on an equal-tempered musical scale that is defined by a reference song, whereby out-of-tune pitches are modified to match the nearest notes on the scale~\cite{autotune, melodyne}.
The naive pitch snapping, however, could lead to modifications lacking lively expressions.
Additionally, pitch errors that deviate much could not be rectified.
Meanwhile, most of the existing solutions limit their scope to pitch correction and ignore rhythmic errors.
All the mentioned problems necessitate extensive manual processes after the modification.

In this paper, we discuss a streamlined framework which automates expressive SVC for both pitch and rhythmic errors under real-world scenarios.
Particularly, the framework is built upon a previous work based on audio synchronization~\cite{luo2018singing}.
This approach regards SVC as a task of dynamic time warping (DTW) followed by vocoding: acoustic features of the input audio (source) and the reference (target) are temporally aligned, whereby the corrected singing voice is synthesized with the pitch contour modified based on the synchronization.
To deal with general sources and targets which might contain musical accompaniments, in this work advanced algorithms for singing voice separation (SVS) and vocal melody extraction are incorporated into the system, which takes a step forward to achieve the real-world application.

In comparison to the previous works~\cite{autotune, melodyne, azarov2014guslar}, the presented method leverages human performances as the target, thereby directly rendering the expressive pitch contours for the correction and avoiding deadpan modifications.
A data-driven model for SVC is proposed recently~\cite{wager2020deep}, which predicts pitch deviations caused by data augmentation, thereby waiving the need of a reference.
However, the model is incapable of fixing large pitch errors as the pitch augmentation used for training is bounded to $\pm{100}$ cents.
Our method, albeit requiring the target, alleviates the demand by the SVS that allows for the leverage of professional renditions which are easily accessible from online streaming services, and is able to rectify pitch errors up to an octave.
Moreover, the derived temporal synchronization allows for correcting rhythmic errors in addition to pitch.
A similar method focuses on transferring singing style~\cite{yong2018} instead of SVC.
On the other hand, we can also achieve style transfer in terms of pitch as a by-product.

\begin{figure}[!t]
    \centering
    \includegraphics[width=\linewidth]{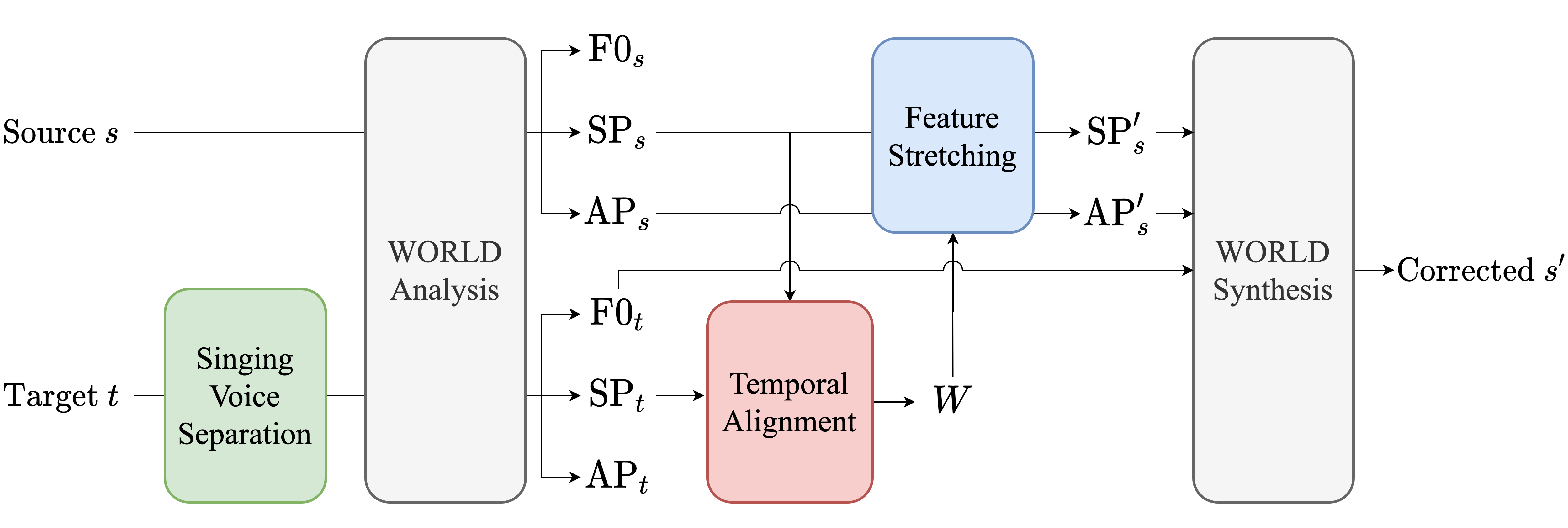}
    \caption{The framework for singing voice correction.}
    \label{fig:system}
\end{figure}

We expect that in the proposed framework, the performance of SVC would benefit from advanced melody extraction techniques.
Therefore, we also conduct a comparative study for state-of-the-art (SoTA) melody extractors.
We then investigate perceptual validity of the standard metrics for melody extraction through the lens of the application of SVC. 
The contributions are summarized as follows:
\begin{itemize}[leftmargin=*]
    \item We present a streamlined framework which automates expressive SVC. 
    It enables both pitch and rhythmic correction, and allows for polyphonic sources and targets, with the aids of SVS and vocal melody extractors.
    Audio samples can be accessed from \texttt{https://yjlolo.github.io/expressive-svc/}.
    \item In conducting the perceptual evaluation, we try to answer, in what aspect the advance of the vocal melody extraction benefits the task of SVC.
    Specifically, we perform correlation analysis for the standard metrics and the subjective judgements, whereby we highlight the strengths and limitations of the existed annotations, metrics, and melody extractors.
\end{itemize}

The remainder of the paper is structured as follows.
We first present the framework in Section~\ref{sec:methods}, and then describe the experimental setups along with the results in Section~\ref{sec:experiment}. 
We have an in-depth discussion in Section~\ref{sec:discussion}, and conclude the study in Section~\ref{sec:conclusion}.

\section{Methods}\label{sec:methods}

As shown in~\figref{fig:system}, the proposed method consists of SVS, acoustic feature extraction, temporal alignment, feature stretching, and audio re-synthesis.
The main architectural differences between the proposed and the previous work~\cite{luo2018singing} are the integration of SVS and the comparative study of SoTA vocal melody extractors.

The inputs to the system are a source $s$ and a target $t$.
In our experiments, $s$ contains only singing voice of the source, while $t$ has singing voice of the target overlapped with background music.
The goal is to correct $s$ in terms of pitch and rhythmic by aligning and rendering parameters from $t$, while keeping vocal timbre of $s$.

Note that although monophonic $s$ is not a requirement for our method, we believe that using monophonic sources and polyphonic targets matches well with the real-world use cases for SVC.

\subsection{Singing Voice Separation}\label{subsec:svs}
We integrate SVS to extract the target singing from the professional renditions $t$ that are easily accessible through online streaming services.
This facilitates the later stages of feature extraction and temporal alignment for singing voice.
We consider SoTA models, namely \textit{Open-Unmix}~\cite{openunmix} and \textit{Demucs}~\cite{demucs} for the task. In the rest of the article, we denote $t$ as the extracted target singing voice.

\subsection{Acoustic Feature Extraction}\label{subsec:feat_extract}
Given an input audio resampled to 16,000Hz, we employ WORLD vocoder~\cite{morise2016world} to extract parameters using short-time Fourier transform with a window size of 64ms and a hop size of 5ms, including pitch contour \txtsub{F0}{}, spectral envelope \txtsub{SP}{}, and aperiodicity \txtsub{AP}{}.
Frame-wise 24-ordered Mel-cepstral coefficients (MCEPs) are then derived from \txtsub{SP}{} which serve as the feature for aligning $s$ and $t$.

Because the framework directly replaces \txtsub{F0}{s} with \txtsub{F0}{t} after the alignment to achieve SVC, we conduct a comparative study for SoTA vocal melody extractors in order to study the relation between accuracy of the extracted F0 and the consequent performance of SVC. Particularly, we include CREPE~\cite{crepe}, PATCH~\cite{patch}, and SEG~\cite{seg} in our study.

\subsection{Singing Voice Alignment}\label{sec:align}

Following the previous work~\cite{luo2018singing}, we use canonical time warping (CTW)~\cite{zhou2009canonical} to temporally align $s$ and $t$.
CTW iteratively projects the input sequences to a common subspace by canonical correlation analysis, and performs DTW therein.
This can be seen as endowing DTW with a feature selection mechanism.


\subsection{Feature Stretching and Audio Re-synthesis}\label{sec:feat_stretch_and_audio_resyn}

As illustrated in \figref{fig:system},
the modified source performance $s'$ is re-synthesized by WORLD,
which consumes the speech parameters \txtsub{F0}{t}, \ptxtsub{SP}{s}, and \ptxtsub{AP}{s};
where \ptxtsub{SP}{s} and \ptxtsub{AP}{s} are obtained by temporal stretching of \txtsub{SP}{s} and \txtsub{AP}{s} given 
the optimal path of alignment $W$ derived from CTW.

In particular, let $f$ denote collectively \txtsub{SP}{} and \txtsub{AP}{}. 
Given temporal associations that align $f_s$ and $f_t$, we either locally average the values of $f_s$ or interpolate them with constants.
This makes the sequence length of the modified feature $f'_s$ equal to that of $f_t$, thereby correcting the rhythmic errors.
Note that we do not directly replace $f_s$ with $f_t$ as we do for \txtsub{F0}{s}, which is to preserve the vocal timbre of $s$ represented by \txtsub{SP}{s}. More advanced manipulations for the feature stretching are left for future investigations. 
\begin{figure*}[!htb]
    \centering
    \includegraphics[trim=0cm 2.1cm 0cm 0cm, clip=true, width=\textwidth]{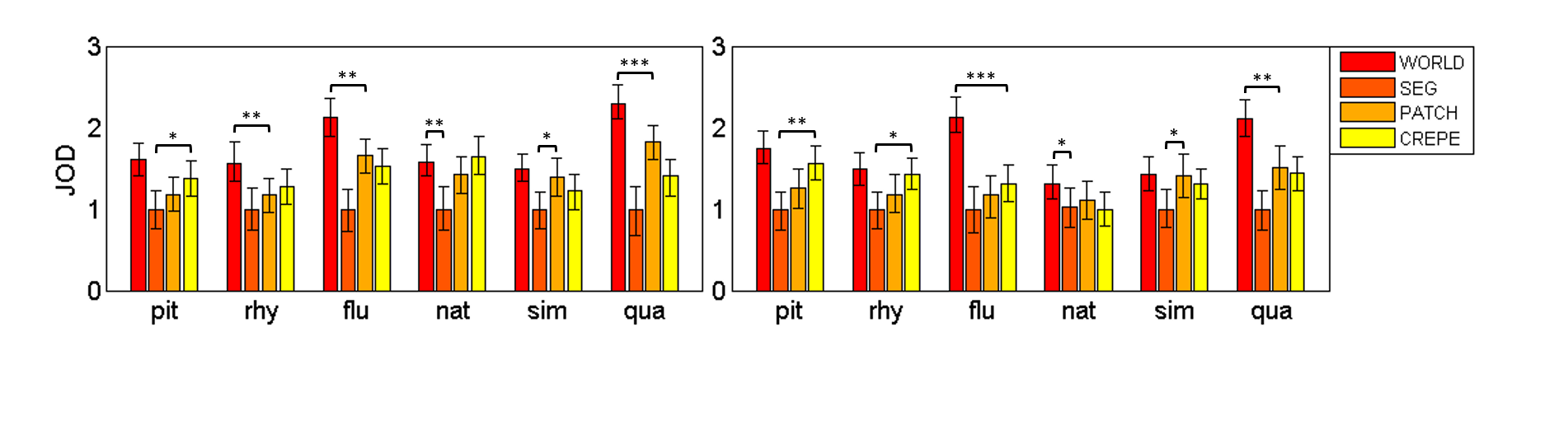}
    \caption{The quality scores measured in JODs for the six criteria. Different numbers of asterisks denote increasing levels of statistical difference, i.e., $p<0.05$, $p<0.01$, and $p<0.001$, respectively. Left: methods based on Open-Unmix for SVS. Right: Demucs for SVS.}
    \label{fig:JOD}
\end{figure*}

\begin{figure*}
    \centering
    \includegraphics[width=\textwidth]{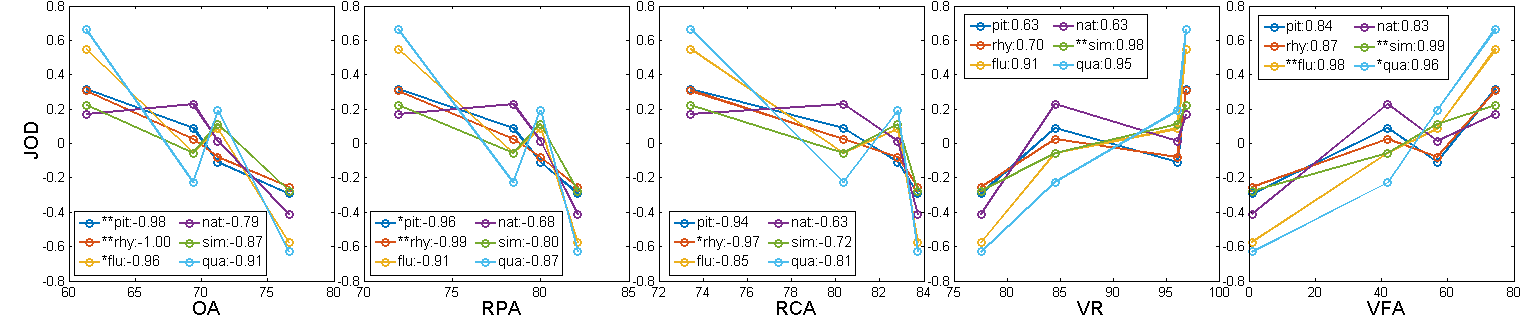}
    \caption{Pearson correlation between the standard metrics and the criteria. Correlation coefficients are shown in the legend. Different numbers of asterisks that precede the criteria denote increasing levels of statistical difference, i.e., $p<0.05$, $p<0.01$, and $p<0.001$, respectively.}
    \label{fig:pearson}
\end{figure*}

\section{Experiments}\label{sec:experiment}

\subsection{Subjective Evaluation}
We use six amateur recordings of 30-second singing voice from MIR-1K~\cite{hsu2010mir1k} as the sources $s$.
The dataset features 1000 clips of mandarin pop songs, with singing voice separated from background music.
For the targets $t$, we leverage the six songs performed by the original singers, respectively, accessed from \textit{YouTube}.\footnote{\texttt{https://www.youtube.com}}

We conduct an experiment of pairwise comparison with online questionnaires.
From the SVS and melody extraction algorithms mentioned in Section~\ref{subsec:svs} and ~\ref{subsec:feat_extract}, we form eight methods from the two SVS and the four vocal melody extractors.
By dividing the eight methods into two groups based on SVS, we have 12 pairs for each of the six songs ($2 \text{ groups of SVS} \times 6 \text{ pairs formed by the four melody extractors}$), which results in 72 pairs in total ($12 \text{ pairs} \times 6 \text{ source songs}$).
We do not exhaust all possible pairs by considering inter-group (SVS) combinations, which is in order to reduce the number of pairs involved in the experiment, thereby shortening the survey and avoiding fatigue.
The 72 pairs of comparison are distributed uniformly in 12 questionnaires, each has six pairs to evaluate corresponding to the six sources.

A subject is asked to first listen to a source, and is then presented a pair of corrected sources by different methods.
The task is to select one that is better than the other, in terms of six criteria usually seen in singing contests~\cite{ising}: \textit{pitch} accuracy, \textit{rhythm} accuracy, \textit{fluency}, \textit{naturalness}, vocal timbre \textit{similarity} between the modified and the original source, and overall audio \textit{quality}.
The subject is tasked to repeat the process for the six songs in a session.
We employ a scheme of two-alternative-forced-choice, which is because our pilot study suggests that not providing ``no preference'' encourages the subjects to scrutinize the samples before making decisions.

We gather responses from 115 subjects and scale the pairwise comparison data to obtain quantitative scores.
The scaling approach~\cite{perezortiz2017practical} defines the quality score of a tested method as a one-dimensional Gaussian distribution, with the mean representing the overall quality and the standard deviation addressing variance of subjective judgement.
It employs a maximum-likelihood estimator such that the scaled quality scores best explain the collected data whose likelihood function is modeled as a binomial distribution:
\begin{equation}
      L(\hat{q}_i - \hat{q}_j | c_{ij}, n_{ij}) = \binom{n_{ij}}{c_{ij}} P(r_i > r_j)^{c_{ij}} (1 - P(r_i > r_j))^{n_{ij} - c_{ij}},
\end{equation}
where $\hat{q}_i$ and $\hat{q}_j$ denote the estimated quality scores of the tested methods $i$ and $j$, respectively; $r_i$ and $r_j$ refer to the random variables sampled from the underlying Gaussian distributions.
$c_{ij}$ denotes the number of responses that $i$ is selected as being better than $j$, and $n_{ij}$ is the total number of trials that $i$ and $j$ are compared. 
We refer readers to the literature~\cite{perezortiz2017practical} for more details.

The quality scores estimated this way are reported to have a strong linear relation with mean opinion score (MOS) obtained by direct rating~\cite{zerman2018}.
Conducting the pairwise comparison is nevertheless easier for non-expert raters, more sensitive to subtle but noticeable quality differences, and invariant to inter-rater scale difference~\cite{shah2015}.
Confidence intervals for the scaled quality scores can be computed using bootstrapping techniques, which gives the $95\%$ confidence intervals for the estimated quality scores.

\begin{table}[!t]
\centering
\scriptsize
\begin{tabular}{cl|ccc|cc}
\toprule
\multicolumn{2}{c|}{Method}          & OA    & RPA   & RCA   & VR    & VFA   \\ \midrule\midrule
\multirow{4}{*}{\begin{tabular}[c]{@{}c@{}}Open-\\ Unmix\end{tabular}}
                            & WORLD & 61.29 & 71.93 & 73.43 & \textbf{96.89} & 74.46 \\
                            & SEG   & \textbf{76.61} & \textbf{82.09} & \textbf{83.72} & 77.59 & \textbf{0.99} \\
                            & PATCH & 71.22 & 79.99 & 82.82 & 96.09 & 56.79 \\
                            & CREPE & 69.38 & 78.45 & 80.35 & 84.53 & 41.92 \\ \bottomrule
\end{tabular}

\caption{Evaluation of the standard metrics for melody extraction.}

\label{tab:mir_eval}
\end{table}

\subsection{Objective Evaluation}
In order to investigate the interplay between the extracted pitch contours and the perceptual study, we also report standard metrics for melody extraction implemented by \texttt{mir\_eval}~\cite{mireval}, including raw pitch accuracy (RPA), raw chroma accuracy (RCA), and overall accuracy (OA) for pitch; and voice recall (VR) and voice false alarm (VFA) for voicing~\cite{salamon2014}.

In particular, the eight methods that we use for the subjective experiment are evaluated with the 1000 clips of music mixtures from MIR-1K.
For the methods based on deep learning, we use pre-trained models provided from the literature.
The averaged scores over individual clips are reported for all the metrics.
\begin{figure*}[!t]
    \centering
    \includegraphics[width=\linewidth]{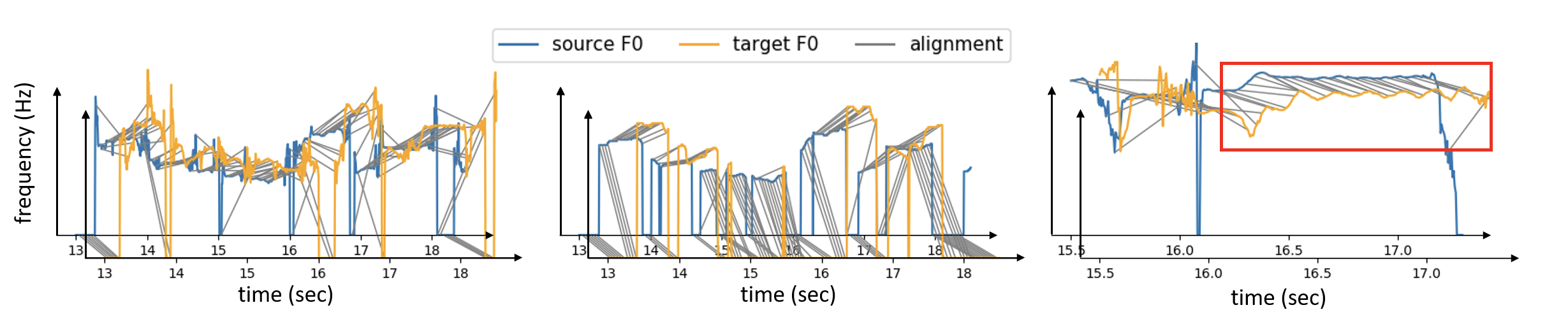}
    \caption{Illustration of the pitch contours and the alignment. Left: WORLD as the vocal melody extractor. Center: SEG as the vocal melody extractor. Right: the red rectangle highlights the alignment from a plain singing (source) to an expressive one (target).}
    \label{fig:alignment}
\end{figure*}

\begin{figure}
    \centering
    \includegraphics[width=0.85\linewidth]{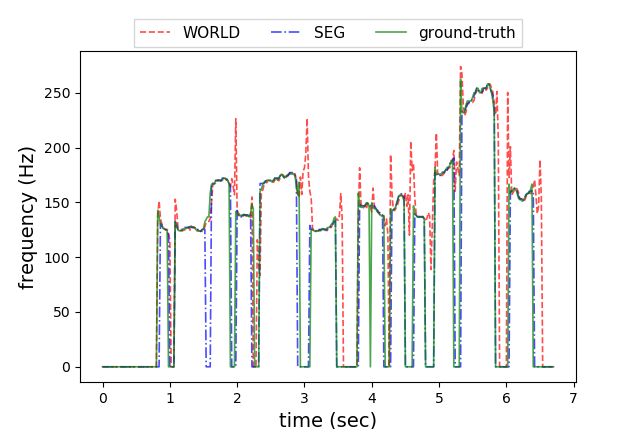}
    \caption{Pitch contours extracted by WORLD and SEG.}
    \label{fig:wvs}
\end{figure}

\subsection{Results}
\figref{fig:JOD} shows results of the subjective evaluation, where the left and the right figures correspond to SVS being Open-Unmix and Demucs, respectively.
The scaled values of the pairwise comparison data are measured in Just-Objectionable-Distances (JODs), which quantify the relative quality difference among the evaluated methods~\cite{perezortiz2017practical}.
Note that the scaling is performed separately for the two groups of SVS, and individually for each of the criteria.
As a result, we can only meaningfully compare within-criteria JODs, and the common x- and y-axes across the criteria are solely for visualization.
Also, the minimum JOD from each of the criteria is set to one, which does not affect interpretability because we focus on the relative difference.

We summarize major findings as that WORLD outperforms SEG significantly across all criteria, and that WORLD is the best-performing method in terms of \textit{fluency} and \textit{quality}.
We will extend the discussion in Section~\ref{sec:discussion}.
Due to the space constraint, we carry on our analysis with Open-Unmix for SVS, as our pilot subjective study with a professional singer indicates superiority of the methods based on Open-Unmix over that based on Demucs.

As shown in~\tabref{tab:mir_eval}, methods based on SEG outperform other melody extractors in terms of OA, RPA, RCA, and VFA; while WORLD prevails only in terms of VR.
Interestingly, the results are at odds with our findings in the subjective evaluation where WOLRD clearly outperforms SEG. 
We measure Pearson correlation to gain more insights, and report in~\figref{fig:pearson} the correlation coefficients calculated from the JODs and the objective scores.
In general, the correlation coefficients are positive for VR and VFA, and are negative for OA, RPA and RCA.
Surprisingly, both of the criteria \textit{pitch} and \textit{rhythm} negatively correlate with OA and RPA.

Before continuing our discussion on the relation between the subjective and the objective evaluation, we demonstrate that the proposed framework is able to render target singing style as illustrated in the right panel of~\figref{fig:alignment}.
This can be seen as a by-product of SVC. 
As addressed by a professional singer during our pilot study, the corrected source is reminiscent of the original artist in terms of articulation and particular style of vibrato.

\section{Discussion}\label{sec:discussion}
We first note that the melody extractor of WORLD, Harvest~\cite{harvest2017}, is able to estimate frequency values in an arbitrary fine resolution (continuous outputs), and is tuned to avoid wrongly estimating voiced frames as unvoiced (thus the high VR and correspondingly the high VFA).
These characteristics are claimed to be critical in high-quality speech synthesis~\cite{harvest2017} which are relevant to the subjective evaluation.
The striking difference can be observed from the left and the center panel in~\figref{fig:alignment}.
It is evidenced that frames of SEG-based contours are much more likely to be predicted as unvoiced, especially seen in the target, which could dramatically hinder the synthesis quality when the frames are actually voiced~\cite{thomas2018}.
Meanwhile, the high VFA of WORLD does not seem to bring degradation to perceptual quality, which is also shown in the previous works~\cite{harvest2017, thomas2018}.
Moreover, as the framework re-synthesizes the corrected source with the target pitch contour, the predicted unvoiced frames are harmful to the quality.
These observations could explain the positive correlation between VR/VFA and JODs, and that WORLD prevails in terms of \textit{quality}.

On the other hand, the SoTA melody extractors such as SEG cast the task of pitch extraction as a problem of classification (discrete outputs), and are trained to optimize the standard metrics which are calculated based on human annotations of pitch contours.
The contrast of WORLD and SEG is further demonstrated in~\figref{fig:wvs}.
Once again, one can spot more gaps between segments along the SEG-based contour.
This actually aligns with the ground-truth and implies the limitations of annotation that will be addressed later.
The discontinuity, together with the discrete outputs, might explain how SoTA models are outperformed by WORLD in terms of \textit{fluency}.
We can also observe ``pitch spikes'' along the WORLD-based contour, which might signify consonants and could be the main culprit that causes the inferior OA, RPA, and RCA.
Modeling the consonants instead of suppressing them as unvoiced, however, improves quality of synthesis as the latter hinders the estimation of \txtsub{SP}{} and \txtsub{AP}{}.



Additionally, WORLD is shown to outperform SEG in terms of \textit{naturalness} in~\figref{fig:JOD}, and the criteria is shown to have negative correlation coefficients with OA, RPA, and RCA; and positive ones with VR and VFA from~\figref{fig:pearson}.
This implies that an accurate pitch contour, evaluated by the existed metrics, is not necessarily natural in perception.
\textit{Similarity} is positively correlated with VR and VFA, which implies that the artifacts caused by the discontinuity of pitch contours could hinder the perception of vocal similarity.

Based on the discussion, we summarize the lessons that we could learn as follows. 
(1) \textit{Limitations of human annotations}: Annotations of pitch contours are hardly agreed by different human annotators.
Time instants associated with ambiguous pitch values further aggravate the problem, such as boundaries of pitched segments and voiced consonants.
On singing voice, a particular methodology is introduced to note-level transcription~\cite{mora2010, Molina2014EvaluationFF}, but is not yet a practice for the pitch contour estimation~\cite{salamon2014}.
An established annotation protocol which precisely accommodates the ambiguity is therefore necessary in order to perform a fairer benchmark with both continuously and discretely valued pitch estimators.
(2) \textit{Limitations of the standard metrics}:
The tolerance of $\pm{50}$ cents for estimation errors is usually used as a workaround for the subjective nature of the annotation. 
An improved metric thus can be the one considering different tolerance levels by e.g. weighted average with weights inversely proportional to the values of thresholds.
A similar idea is proposed for the voicing estimation~\cite{Bittner2019GeneralizedMF}.
Moreover, according to our experimental results,
continuity of the pitch contour may play an important role in perceptual evaluation, which is also taken into account by previous works~\cite{Frieler2019, Bosch2016}.
(3) \textit{Limitations of classification-based melody extractors}: Discrete predictions of pitch values may lead to contours lacking in naturalness and fluency, which are detrimental to synthesis.
They are however favored by the standard metrics.
On the other hand, the F0 estimator that produces continuous outputs, albeit inferior in terms of the metrics due to the pitch variations nearby pitched segments, is favored perceptually.
This again hints towards the lack of the generalizability of the existed metrics~\cite{salamon2019}.

\section{Conclusion}\label{sec:conclusion}

We devise a streamlined framework for automatic SVC which accommodates real-world use cases, correct both pitch and rhythmic errors, and render expressive parameters from professional singers.
We investigate the perceptual validity of the standard metrics for vocal melody extraction through the lens of the application of SVC.
The results show that a good performance in terms of most of the metrics does not signify better perceptual scores, whereby we highlight the challenges posed to the existed annotations, metrics, and melody extractors.
Paths forward involve devising a robust quantitative metric which accounts for the issues presented in the discussion.

\bibliographystyle{IEEEbib}
\bibliography{strings,refs}

\end{document}